\documentclass[12pt]{article}
\usepackage{epsf}
\newcommand{\pl}{Phys.\ Lett.}
\newcommand{\np}{Nucl.\ Phys.}

\title{ {\bf
$B_s\rightarrow \gamma \gamma$ decay in model III and 3HDM($O_2$)
with $CP$ violating effects.}}
\author{\vspace{1cm}\\
         {\bf M. Boz} 
\thanks{E-mail address:
        mugeboz@thep1.hun.edu.tr}
 \\
        Physics Department, Hacettepe University \\
        Ankara, Turkey\\
\vspace{5mm}\\
        {\bf E. O. Iltan}
        \thanks{E-mail address:
        eiltan@heraklit.physics.metu.edu.tr}
 \\
        Physics Department, Middle East Technical University \\
        Ankara, Turkey\\}
\date{}
\begin{document}
\setlength{\baselineskip}{24pt}
\maketitle
\setlength{\baselineskip}{7mm}
\begin{abstract}
We analyse the $CP$ asymmetry for $B_{s}\rightarrow\gamma\gamma$ in the two 
Higgs doublet model with tree level flavor changing currents (model III) and 
three Higgs doublet model with $O_2$ symmetry in the Higgs sector, including
$O_{7}$ type long distance effects. Further, we study the dependencies of 
the branching ratio $Br(B_{s}\rightarrow\gamma\gamma)$ and the ratio of 
$CP$-even and $CP$-odd amplitude squares, $R=|A^{+}|^2/|A^{-}|^2$, on the 
$CP$ parameter $sin\,\theta$. We found that, there is a weak $CP$ asymmetry, 
at the order of $10^{-4}$. Besides, the branching ratio 
$Br(B_{s}\rightarrow\gamma\gamma)$, and also $R$ ratio, is not sensitive to 
the $CP$ parameter for $|\frac{\bar{\xi}^{U}_{N,tt}}{\bar{\xi}^{D}_{N,bb}}|<1$.
\end{abstract}
\thispagestyle{empty}
\newpage
\setcounter{page}{1}
\section{Introduction}
One of the most important classes of decays are rare B-meson decays
which are induced by flavor changing neutral currents (FCNC) at loop
level in the Standard Model (SM). Therefore, one can obtain
quantitative  information for the SM parameters , such as 
Cabbibo-Kobayashi-Maskawa (CKM) matrix elements, leptonic decay constants, 
$CP$ ratio, etc. and  the  measurement of physical quantities like branching
ratio ($Br$), $CP$ asymmetry $(A_{CP})$,{\ldots} etc, gives important
clues about the model under consideration. These decays  are also
sensitive to the  physics beyond the SM, such as
two Higgs Doublet model (2HDM), Minimal Supersymmetric extension of the SM
(MSSM) \cite{Hewett}.
 
$B_s\rightarrow \gamma\gamma$ decay, which is an example of the
rare B meson decays,  is an important candidate to test the theoretical
models and to construct new models  in the framework of the 
planned experiments at the upcoming KEK and SLAC-B factories and existing
hadronic accelerators. $B_s\rightarrow \gamma\gamma$ decay, induced by the 
process $b\rightarrow s \gamma\gamma$ in the quark level, has been studied 
in the SM  \cite{kalinowski}-\cite{simma} and 2HDM \cite{aliev} without QCD 
corrections. Since the QCD corrections to the inclusive decay $b\rightarrow
s\gamma$ are large (see \cite{Grinstein} - \cite{burasmisiak} and references 
therein), it is expected that they are also large in the inclusive  
$b\rightarrow s\gamma\gamma$ decay and, therefore, in the exclusive 
$B_s\rightarrow \gamma\gamma$ decay. With  the addition of 
the leading logarithmic (LLog) QCD corrections, the analysis has been 
repeated in the SM \cite{Gud}-\cite{Soni}, 2HDM \cite{gudalil},  
MSSM \cite{bertolini} , and the strong sensitivity to these corrections was 
obtained. Recently, $B_s\rightarrow \gamma\gamma$ decay has been  calculated 
in the framework of model III \cite{alil4} with the addition of the LLog QCD
corrections and the current upper limit of the $ Br$ ratio 
$Br (B_s\rightarrow \gamma\gamma) \le 1.48\, . 10^{-4}$ \cite{Acciarri}  
was theoretically obtained for larger values of the Yukawa coupling 
$\bar{\xi}_{N,bb}$ and the ratio $|r_{tb}|=
|\frac{\bar{\xi}_{N,tt}}{\bar{\xi}_{N,bb}}| >> 1$ (see Appendix $C$ for
definitions).

In the present work, we study $B_s\rightarrow \gamma\gamma$ decay in
2HDM (model III)  and
three Higgs doublet model with $O_{2}$ symmetry in the Higgs sector
$(3HDM(O_{2}))$ \cite{eril5}, with complex Yukawa couplings.
In our calculations, we take into account
the perturbative QCD corrections in the LLog approximation
by following  a method based on heavy quark effective theory (HQET) for 
the bound state of the $B_s$ meson \cite{Gud} and we also consider
the  long-distance effects due to the process $
B_s \to \phi \gamma \to \gamma \gamma$, \cite{Gud}. Since
the complex Yukawa couplings are chosen in the calculations,
in addition to the CKM matrix elements, there  is a new source for $CP$
violation. Using this new source, we obtain  $A_{CP}$ in model III
at the order of $10^{-4}$, which is a small effect.
Furthermore, we calculate this quantity in the $3HDM(O_{2})$
and find that  there is a small decrease as compared to the former
one for $\sin\,\theta \leq 0.5$. Finally, the calculations of $Br$ and the 
$CP$ ratio $R=|\frac{A^{+}}{A^{-}}| $ show that these physical quantities 
are not sensitive to the models under consideration for
$|r_{tb}|=|\frac{\bar{\xi}^{U}_{N,tt}}{\bar{\xi}^{D}_{N,bb}}|<1$
(see Section 3).

The paper is organized as follows:
In section 2,  we present the LLog QCD corrected amplitude for the exclusive
decay $B_s\rightarrow \gamma\gamma$ and the $CP$-even $A^+$ and 
$CP$-odd $A^-$ amplitudes in a HQET inspired approach. 
Then, we calculate $A_{CP}$, assuming that the Yukawa couplings 
are complex in general and also derive $Br$. Finally, these calculations 
are repeated in the $3HDM(O_{2})$ model with the redefinition of the 
Yukawa combination $\lambda_{\theta}$ (see eq. (\ref{l3})). Section 3 is 
devoted to our analysis of the  physical quantities under consideration 
and our conclusions. In the Appendix, we give a brief explanation about 
the model III and $3HDM(O_2)$ which we study. Further, we present the 
operator basis and the Wilson coefficients responsible for the inclusive 
$b\rightarrow s\gamma\gamma$ decay in the model III. Finally, we give the 
explicit forms of some functions appearing in the Wilson coefficients.
\section{\bf Leading logarithmic improved short-distance contributions in 
the model III for the decay $B_s\rightarrow \gamma \gamma $ }
The LLog corrected  effective Hamiltonian in the model III (see Appendix $A$) 
for the exclusive $B_s\rightarrow  \gamma \gamma $ decay is 
\begin{eqnarray}
{\cal{H}}_{eff}=-4 \frac{G_{F}}{\sqrt{2}} V_{tb} V^{*}_{ts} 
\sum_{i}\left(C_{i}(\mu) O_{i}(\mu) + C'_{i}(\mu) O'_{i}(\mu)\right) \,
 \, , 
\label{hamilton}
\end{eqnarray}
where the $O_{i}$, $O'_{i}$ are operators given in eqs.~(\ref{op1}),
~(\ref{op2}), and $C_{i}$, $C'_{i}$ are the Wilson coefficients renormalized 
at the scale $\mu$. The Wilson coefficients $C_{i}$ and  $C'_{i}$ can be 
calculated perturbatively and their explicit forms of at
$m_W$ level are presented in Appendix $B$. The effective Hamiltonian 
(\ref{hamilton}) is obtained by integrating out the heavy degrees of freedom, 
i.e , $t$ quark, $W^{\pm}, H^{\pm}, H_{1}$, and $H_{2}$ bosons in the present 
case where $H^{\pm}$ denote charged, $H_{1}$ and $H_{2}$ denote neutral 
Higgs bosons. Further, QCD corrections are done through matching the full 
theory with the effective low energy theory at the high scale $\mu=m_{W}$ 
and evaluating the Wilson coefficients from $m_{W}$ down to the lower scale 
$\mu\sim O(m_{b})$. In our case,  we choose the higher scale as $\mu=m_{W}$ 
since the evaluation from the scale $\mu=m_{H^{\pm}}$ to $\mu=m_{W}$ gives 
negligible contribution to the Wilson coefficients ($\sim 5\%$) since the 
charged Higgs boson is heavy due to the current theoretical restrictions 
(see \cite{gudalil}, \cite{ciuchini} )

The decay   amplitude for the $B_s\rightarrow \gamma\gamma$
decay is obtained by  sandwiching  the effective Hamiltonian between
the $B_s$ and two photon states, i.e. $<B_s| {\cal{H}}_{eff}|\gamma\gamma>$, 
and the  matrix element can be written in terms of two Lorentz structures 
\cite{simma} - \cite{aliev}, \cite{Gud}-\cite{Soni}:  
\begin{equation}
{\cal A}(B_{s}\rightarrow \gamma \gamma)=
A^{+} {\cal F}_{\mu\nu} {\cal F}^{\mu\nu} +
i A^{-} {\cal F}_{\mu\nu} \tilde{{\cal F}}^{\mu\nu}
\, \, ,
\label{amp}
\end{equation}
where 
$\tilde{{\cal F}}_{\mu\nu}=\frac{1}{2}\epsilon_{\mu\nu\alpha\beta} 
{\cal F}^{\alpha\beta}$ and $A^{+}$ ($A^{-}$) is the $CP$-even ($CP$-odd)
amplitude. Denoting $CP$-even ($CP$-odd) amplitude coming from the 
first operator set eq. (\ref{op1}) as $A_1^{+}$ ($A_1^{-}$) and the primed 
operator set eq. (\ref{op2}) as $A_2^{+}$ ($A_2^{-}$), in a HQET inspired 
approach, we get,
\begin{eqnarray}
A_1^{+}&=&\frac{\alpha_{em} G_F}{\sqrt{2} \pi} \frac{f_{B_s}}{m_{B_{s}}^{2}} 
\lambda_t 
\left( \frac{1}{3}
\frac{m^4_{B_s} (m_b^{eff}-m_s^{eff})}{\bar{\Lambda}_s 
(m_{B_s}-\bar{\Lambda}_s) (m_b^{eff}+m_s^{eff})} 
C_7^{eff}(\mu)
\nonumber
\right.\\
&-&
\left.
\frac{4}{9} \frac{m_{B_{s}^2}}{m_b^{eff}+m_s^{eff}}\, [ \,
(-m_b J(m_b)+ m_s J(m_s) ) D(\mu) - m_c J(m_c)  E(\mu)\, ]\,  
\right)  ,\nonumber\\
A_1^{-}&=&- \frac{\alpha_{em} G_F}{\sqrt{2} \pi} f_{B_s} \lambda_t 
\left( \frac{1}{3}
\frac{1}{m_{B_s} \bar{\Lambda}_s (m_{B_s}-\bar{\Lambda}_s)} g_{-}
C_7^{eff}(\mu) -\sum_q Q_q^2 I(m_q) C_q(\mu) \nonumber 
\right.\\
&+&
\left.
\frac{1}{9 (m_b^{eff}+m_s^{eff})} 
[\, (m_b \triangle(m_b)+m_s \triangle(m_s)) D(\mu)+ m_c \triangle(m_c)
E(\mu) \,] \right) \, \, ,
\label{Amplitudes1}
\end{eqnarray}
and, 
\begin{eqnarray}
A_2^{+}&=&\frac{\alpha_{em} G_F}{\sqrt{2} \pi} \frac{f_{B_s}}{m_{B_{s}}^{2}} 
\lambda_t 
\left( \frac{1}{3}
\frac{m^4_{B_s} (m_b^{eff}-m_s^{eff})}{\bar{\Lambda}_s 
(m_{B_s}-\bar{\Lambda}_s) (m_b^{eff}+m_s^{eff})} 
C_7^{\prime eff}(\mu) \right)   ,\nonumber \\
A_2^{-}&=&- \frac{\alpha_{em} G_F}{\sqrt{2} \pi} f_{B_s} \lambda_t 
\left( \frac{1}{3}
\frac{1}{m_{B_s} \bar{\Lambda}_s (m_{B_s}-\bar{\Lambda}_s)} g_{-}
C_7^{\prime eff}(\mu)  \right) \, \, ,
\label{Amplitudes2}
\end{eqnarray}
However, we do not take the $CP$-even and $CP$-odd amplitudes corresponding 
to the primed operator set since their contribution is small as compared to 
former ones (see \cite{alil4,alil}). In eqs. (\ref{Amplitudes1}) and 
(\ref{Amplitudes2}), $Q_q=\frac{2}{3}$ for $q=u,c$ and $Q_q=-\frac{1}{3}$ 
for $q=d,s,b$. Here, we have  used the unitarity of the CKM-matrix 
$\sum_{i=u,c,t} V_{is}^{*} V_{ib}=0 $,  and also the contribution due to 
$V_{us}^{*} V_{ub} \ll V_{ts}^{*} V_{tb}\equiv \lambda_t$ is neglected. The 
function $g_{-}$ is defined as \cite{Gud}:
\begin{eqnarray}
g_{-}=m_{B_s}(m_b^{eff}+m_s^{eff})^2+
\bar{\Lambda}_s (m^2_{B_s}-(m_b^{eff}+m_s^{eff})^2) 
\,\, .
\label{gmin}
\end{eqnarray}
The parameter $\bar{\Lambda}_{s}$ enters in eqs. (\ref{Amplitudes1})
and (\ref{Amplitudes2}) through the bound state kinematics 
(for details see \cite{Gud}). In the expression (\ref{Amplitudes1}), 
the LLog QCD corrected Wilson coefficients $C_{1 \dots 10}(\mu)$ 
\cite{Gud} - \cite{Soni} are,
\begin{eqnarray}
C_u(\mu)&=&C_d(\mu)=(C_3(\mu)-C_5(\mu)) N_c +C_4(\mu)-C_6(\mu) \, \, , 
\nonumber \\
C_c(\mu)&=&
(C_1(\mu)+C_3(\mu)-C_5(\mu)) N_c +C_2(\mu)+C_4(\mu)-C_6(\mu) \, \, ,
\nonumber \\
C_s(\mu)&=&C_b(\mu)=
(C_3(\mu)+C_4(\mu))(N_c+1)-N_c C_5(\mu)-C_6(\mu) \, \, , \nonumber \\
D(\mu)&=&C_5(\mu)+C_6(\mu) N_c \, \, , \nonumber \\
E(\mu)&=&C_{10}(\mu)+C_9(\mu) N_c
\end{eqnarray} 
and  the effective coefficient $C_7^{eff}(\mu)$, defined in the NDR scheme, 
is \cite{alil}
\begin{eqnarray}
C_{7}^{eff}(\mu)&=&C_{7}^{2HDM}(\mu)+ Q_d \, 
(C_{5}^{2HDM}(\mu) + N_c \, C_{6}^{2HDM}(\mu))\nonumber \, \, , \\
&+& Q_u\, (\frac{m_c}{m_b}\, C_{10}^{2HDM}(\mu) + N_c \, 
\frac{m_c}{m_b}\,C_{9}^{2HDM}(\mu))
\label{C7eff}
\end{eqnarray}
where $N_{c}$ is the number of colours ($N_{c}=3$\, for QCD).
The functions $I(m_q)$, $J({m_q})$ and $\triangle 
(m_q)$  come from the irreducible diagrams with an internal $q$ type quark 
propagating and their explicit forms  are given in Appendix $D$. 
In our numerical analysis we used the input values given in  
Table~(\ref{input}).
\begin{table}[h]
        \begin{center}
        \begin{tabular}{|l|l|}
        \hline
        \multicolumn{1}{|c|}{Parameter} & 
                \multicolumn{1}{|c|}{Value}     \\
        \hline \hline
        $m_c$                   & $1.4$ (GeV) \\
        $m_b$                   & $4.8$ (GeV) \\
        $\alpha_{em}^{-1}$      & 129           \\
        $\lambda_t$            & 0.04 \\
        $\Gamma_{tot}(B_s)$             & $4.09 \cdot 10^{-13}$ (GeV)   \\
        $f_{B_s}$             & $0.2$ (GeV)  \\   
        $m_{B_s}$             & $5.369$ (GeV) \\
        $m_{t}$             & $175$ (GeV) \\
        $m_{W}$             & $80.26$ (GeV) \\
        $m_{Z}$             & $91.19$ (GeV) \\
        $\Lambda^{(5)}_{QCD}$             & $0.214$ (GeV) \\
        $\alpha_{s}(m_Z)$             & $0.117$  \\
        $\lambda_2$             & $0.12$ $(\mbox{GeV}^2)$ \\
        $\lambda_1$             & $-0.29$ $(\mbox{GeV}^2)$ \\ 
        $\bar{\Lambda}_s$             & $590$ $(\mbox{MeV})$ \\
        $\bar{\Lambda}$             & $500$ $(\mbox{MeV})$ \\
        \hline
        \end{tabular}
        \end{center}
\caption{Values of the input parameters used in the numerical
          calculations unless otherwise specified.}
\label{input}
\end{table}

At this stage, we will calculate the $CP$ violating asymmetry for the given 
process. The possible sources of such effects are the complex Yukawa couplings 
in the model III. Here, we neglect all the Yukawa couplings except 
$\bar{\xi}^{U}_{N,tt}$ and $\bar{\xi}^{D}_{N,bb}$ (see Section 3).
Using the expressions for the decay amplitude
\begin{eqnarray}
\Gamma=\frac{1}{32\pi m_{B_{s}}}[4|A^{+}|^{2}+\frac{1}{2}m_{B_s}^{4}|
A^{-}|^{2}]
\label{decamp}
\end{eqnarray}
and the $CP$-asymmetry
\begin{eqnarray}
A_{CP}= \frac{\Gamma(B_{s}\rightarrow \gamma \gamma)-
\Gamma(\bar{B_{s}}\rightarrow \gamma \gamma)} 
{\Gamma(B_{s}\rightarrow\gamma\gamma)
+\Gamma(\bar{B_{s}}\rightarrow\gamma\gamma)}\,\, ,   
\label{Acp1}
\end{eqnarray}
we get
\begin{eqnarray}
A_{CP}= 2 Im(\lambda_{\theta})\, \frac{8 Im(T^{(+)}_{1}T_{2}^{(+)*})+
m^{4}_{B_s}Im(T^{(-)}_{1} T_{2}^{(-)*})}{D}
\label{Acp2}
\end{eqnarray}
where
\begin{eqnarray}
T^{(+)}_{1}&=&aP_{1}\nonumber\\
T^{(+)}_{2}&=&aP_{2}+b\nonumber\\
T^{(-)}_{1}&=&cP_{1}\nonumber\\
T^{(-)}_{2}&=&cP_{2}+d
\label{decamp2}
\end{eqnarray}
and
\begin{eqnarray}
a&=&\frac {\alpha_{em}G_{F}}{\sqrt{2}\pi}  \frac{f_{B_s}}{3}
\frac{m_{B_s}^{2}}{\bar{\Lambda}_{s}} \, \lambda_t 
\frac{(m_{b}^{eff}-m_{s}^{eff})} {(m_{B_s}-\bar{\Lambda}_{s})(m_{b}^{eff}+
m_{s}^{eff})}\nonumber\\ 
b&=&-\frac{\alpha_{em}G_{F}}{\sqrt{2}\pi} 
\frac{4 f_{B_s}}{9 ({m_{b}}^{eff}+{m_{s}}^{eff})} \, \lambda_{t} 
[(-m_{b}J(m_{b})+m_{s}J(m_{s})] D(\mu)-m_{c}J(m_{c}) E(\mu)]\nonumber\\
c&=&-\frac{\alpha_{em}G_{F}}{\sqrt{2}\pi}  f_{B_s}
\frac{1}{3 m_{B} \bar{\Lambda}_{s}(m_{B_s}-\Lambda_{s})}\,  \lambda_t \, 
g_{-} \nonumber\\
d&=&-\frac{\alpha_{em}G_{F}}{\sqrt{2}\pi}  f_{B} \lambda_{t} [\, \sum_{q} 
Q_{q}^{2} I(m_{q}) C_{q}(\mu) +\nonumber\\
&&
\frac{1}{9 (m_{b}^{eff}+m_{s}^{eff})}\{(m_{b}
\Delta(m_{b})+m_{s} \Delta(m_{s})) D(\mu)+ 
m_{c}\Delta(m_{c})E(\mu)\} \,]
\label{decamp3}
\end{eqnarray}
In eq. (\ref{Acp2}), we used the parametrization,
\begin{eqnarray}
C_{7}^{eff}(\mu)=P_1(\mu)\lambda_{\theta}+ P_{2}(\mu)
\label{param1}
\end{eqnarray}
with
\begin{eqnarray}
\lambda_{\theta}=\frac{1}{m_t m_{b}}|\bar{\xi}^{U}_{N,tt}           
\bar{\xi}^{D}_{N,bb}|e^{i\theta}
\label{l2}
\end{eqnarray}
Here, $\bar{\xi}^{U}_{N,tt}$ is chosen to be as  real   and
$\bar{\xi}^{D}_{N,bb}$ as complex, namely
$\bar{\xi}^{D}_{N,bb}=| \bar{\xi}^{D}_{N,bb}|e^{i\theta}$. Finally the
functions $P_{1}(\mu)$ and $P_{2}(\mu)$, in the LLog approximation
\cite{eril4}, are
\begin{eqnarray}
P_{1}(\mu)&=&\eta^{16/23}F_{2}(y)+\frac{8}{3}(\eta^{14/23}-\eta^{16/23})
G_{2}(y)\nonumber\\
P_{2}(\mu)&=&\eta^{16/23}[C_{7}^{SM}(m_{W}+
\frac{|\bar{\xi}^{U}_{N,tt}|^{2}}{m_{t}^{2}}F_{1}(y)]\nonumber\\
&&
+\frac{8}{3}(\eta^{14/23}-\eta^{16/23})
[C_{8}^{SM}(m_{W}+
\frac{|\bar{\xi}^{U}_{N,tt}|^{2}}{m_{t}^{2}}G_{1}(y)]\nonumber\\
&&
+Q_{d}(C_{5}^{LO}(\mu)+N_{c}C_{6}^{LO}(\mu))+
+Q_{u}(\frac{m_{c}}{m_{b}}C_{12}^{LO}(\mu)+
N_{c}\frac{m_{c}}{m_{b}}C_{11}^{LO}(\mu))\nonumber\\
&&
+C_{2}(m_{W}) \sum_{i=1}^{8}{h_{i}\eta^{\alpha_{i}}}
\label{funpp}
\end{eqnarray}
and
\begin{eqnarray}
D&=&|\lambda_{\theta}|^{2}\{8 |T_{1}^{(+)}|^{2}+m_{B_s}^{4}|T_{2}^{(-)}|^{2}\}
+2 Re(\lambda_{\theta})\{8 Re(T^{(+)}_{1}T^{(+)*})\nonumber\\
&& + m^{4}_{B_s}Re(T^{(-)}_{1}T^{(-)*})\}+\{8|T_{2}^{(+)}|^{2}+m_{B_s}^{4}|
T_{2}^{(-)}|^{2}\}
\label{denom}
\end{eqnarray}
In eq. \ref{funpp}, $\eta=\alpha_s (m_W)/\alpha_s (\mu)$ and  $h_i$, $a_i$
are numbers which appear during the evaluation of the Wilson coefficients
\cite{buras2}. 

Now, we would like to add the LD distance contributions due to the process
$B_{s}\rightarrow\phi\gamma \rightarrow\gamma\gamma$ \cite{Gud}. These effects
can be taken into account by the redefinition of the functions  $ T^{(+)}_{1},
T^{(+)}_{2},  T^{(-)}_{1}, T^{(-)}_{2}$,
\begin{eqnarray}
T^{\prime (+)}_{1}&=&T^{(+)}_{1} + P_{1}\, a_{LD}^{(+)}\nonumber\\
T^{\prime (+)}_{2}&=&T^{(+)}_{2} + P_{2}\, a_{LD}^{(+)}\nonumber\\
T^{\prime (-)}_{1}&=&T^{(+)}_{1} + P_{1}\, a_{LD}^{(-)}\nonumber\\
T^{\prime (-)}_{2}&=&T^{(+)}_{2} + P_{2}\, a_{LD}^{(-)}
\label{decampld}
\end{eqnarray}
where 
\begin{eqnarray}
a_{LD}^{(+)}&=&-\sqrt{2}\frac{\alpha_{em}G_{F}}{\pi}
\bar{F_{1}}(0)f_{\phi}(0)\lambda_{t}\frac{m_{b}(m_{Bs}^{2}-m_{\phi}^{2})}
{3 m_{\phi}m_{Bs}^{2}}\nonumber\\
a_{LD}^{(-)}&=&\sqrt{2}\frac{\alpha_{em}G_{F}}{\pi}
\bar{F_{1}}(0)f_{\phi}(0)\lambda_{t}\frac{m_{b}}{3 m_{\phi}}
\label{ldaa}
\end{eqnarray}
Note that there is also a $LD$ contribution due to the chain process 
$B_s \rightarrow \phi\psi \rightarrow \phi\gamma \rightarrow \gamma\gamma$
and it is negligible compared to the one due to the decay 
$B_s \rightarrow \phi\gamma \rightarrow \gamma\gamma$ \cite{gudil}.

Finally, we derive  the branching ratio for the given process as
\begin{eqnarray}
Br&=&\frac{1}{ 64\,\pi\,m_{B_s} \Gamma_{tot}}
[\,|\lambda_{\theta}|^{2} \{8 |T_1^{(+)}|^{2}+m_{B_s}^{4}|T_1^{(-)}|^{2}\}
+4Re(\lambda_{\theta})\{8 T^{(+)}_{1}Re(T^{(+)}_{2}) \nonumber\\
&&
+m^{4}_{B_s}T^{(-)}_{1}Re(T^{(-)}_{2})\}+
\{8|T_{2}^{(+)}|^{2}+m_{B}^{4}|T_{2}^{(-)}|^{2}\}\,] 
\label{brand}
\end{eqnarray}

In the $3HDM(O_{2})$(see Appendix $C$ and  \cite{eril5}), the 
physical quantities under consideration are the same with the 
redefinition of $\lambda_{\theta}$, 
\begin{eqnarray}
\lambda_{\theta}=\frac{1}{m_t m_{b}} \bar{\epsilon}^{U}_{N,tt}
\bar{\epsilon}^{D}_{N,bb} (\cos^{2}\theta+i\sin^{2}\theta)
\label{l3}
\end{eqnarray}
\section{Discussion}
In the model III, there are many parameters, such as complex Yukawa 
couplings, $\xi^{U,D}_{i,j}$ (i,j are flavor indices), masses of charged 
and neutral Higgs bosons and they should be restricted using experimental
results. All the Yukawa couplings except $\xi^{U}_{N,tt}$ and 
$\xi^{D}_{N,bb}$ are cancelled based on the experimental 
measurements by CLEO \cite{cleo}, 
\begin{eqnarray}
Br (B\rightarrow X_s\gamma)= (3.15\pm0.35\pm0.32)\, 10^{-4} \,\, ,
\label{br2}
\end{eqnarray}
$\Delta F=2$ mixing and the $\rho$ parameter \cite{atwood}
(see \cite{alil} for details). This discussion also allows us to neglect 
the contributions coming from the primed operator set. Further, the $CP$ 
parameter $"\theta"$, appearing in the combination
$\bar{\xi}^{U}_{N,tt}\bar{\xi}^{D*}_{N,bb}= |\bar{\xi}^{U}_{N,tt}
{\xi}^{D}_{N,bb}|e^{-i\theta}$,
is restricted due to the experimental upper limit on neutron electric
dipole moment $d_{n}< 10^{-25}$ e.cm, which gives an upper bound to the
combination $\frac{1}{m_t m_{b}} Im( \bar{\xi}^{U}_{N,tt}
\bar{\xi}^{D *}_{N,bb}) <1.0$ for $m_{H}\approx 200 $ Gev \cite{Chao}.

Now, we are ready to start with the discussion of $CP$ asymmetry in our
process. Note that, in the analysis,  $|C_{7}^{eff}|$
is allowed to lie in the region, $ 0.257 \le |C_{7}^{eff}|\le 0.439 $
due to the CLEO measurement (see \cite{alil}) and the parameters
$\bar{\xi}^{U}_{N,tt}$, $\bar{\xi}^{D}_{N,bb}$ and $\theta$ are
restricted. We choose the scale as $\mu=\frac{m_{b}}{2}$ since we predict
that this choice reproduce effectively the Next to Leading Order (NLO) 
result (see \cite{Gud}). 

\begin{figure}[htb]
\vskip -3.0truein
\centering
\epsfxsize=6.8in
\leavevmode\epsffile{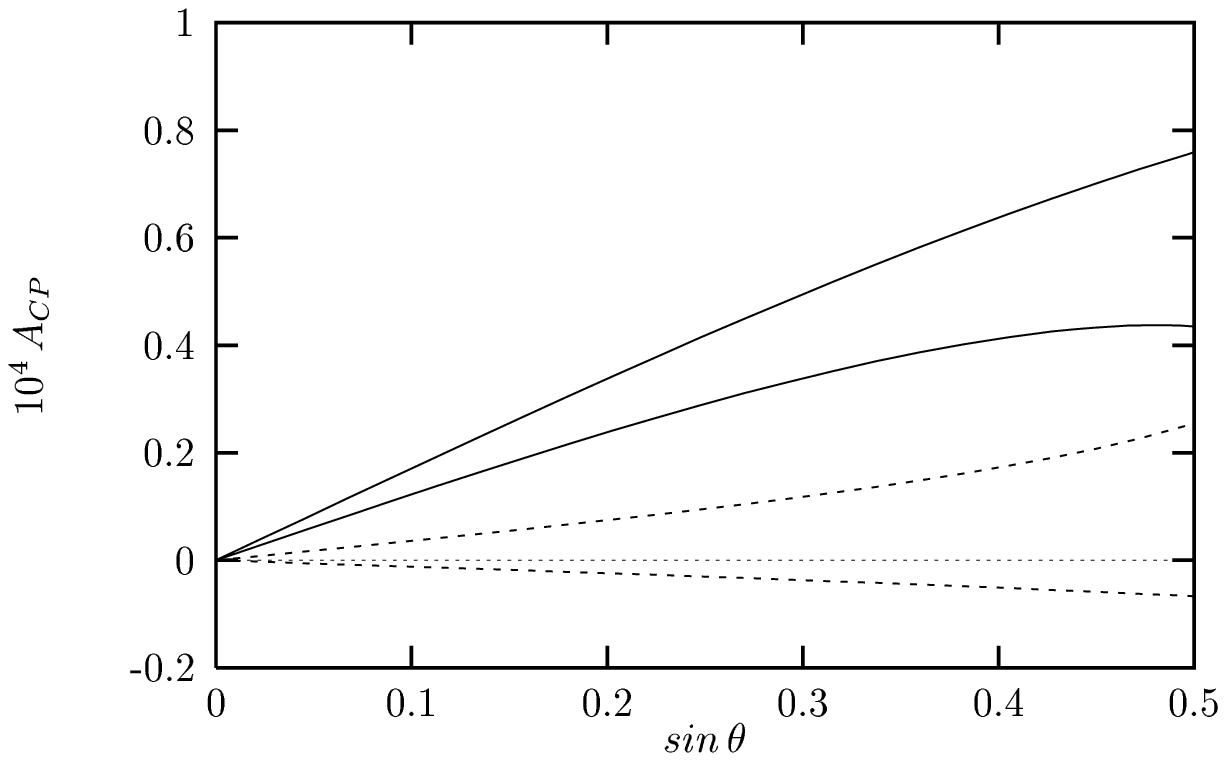}
\vskip -3.0truein
\caption[]{$A_{CP}$ as a function of  $sin\,\theta$ for
$\bar{\xi}_{N,bb}^{D}=40\, m_b$ and $m_H^{\pm}=400\, GeV$ in the region 
$|r_{tb}|<1$, at the scale $\mu=m_b/2$, without LD effects, in model III. 
Here $A_{CP}$ is restricted in the region bounded by solid lines for 
$C_7^{eff}>0$ and by  dashed lines for $C_7^{eff}<0$. }
\label{Acp0s}
\end{figure}

\begin{figure}[htb]
\vskip -3.0truein
\centering
\epsfxsize=6.8in
\leavevmode\epsffile{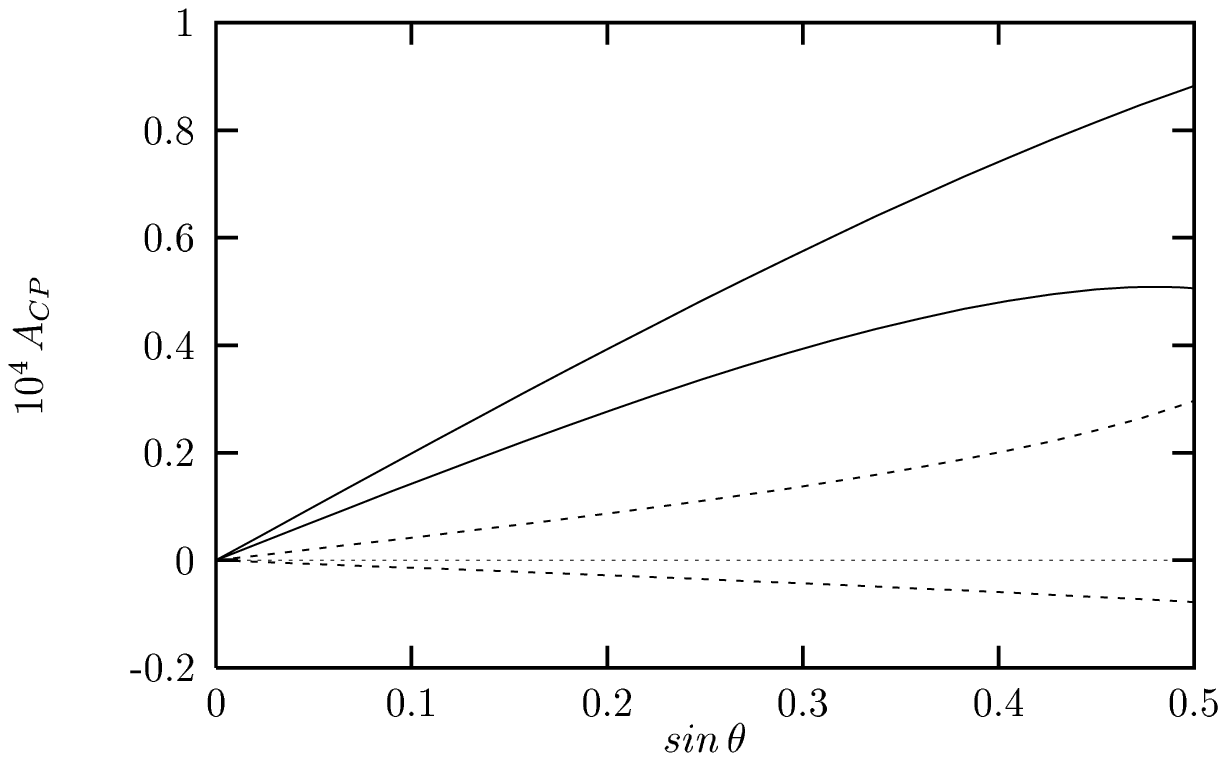}
\vskip -3.0truein
\caption[]{The same as Fig. \ref{Acp0s} but with LD effects.}
\label{Acpps}
\end{figure}

In Fig \ref{Acp0s}(\ref{Acpps}), we plot the $A_{CP}$ of the decay 
$B_{s}\rightarrow \gamma\gamma$ with respect to $\sin\theta$, for 
$\bar{\xi}^{D}_{N,bb}=40 \, m_{b}$ and  $m_{H^{\pm}}=400$ GeV, in the 
case where the ratio 
$|r_{tb}|=|\frac{\bar{\xi}_{N,tt}^{U}}{\bar{\xi}_{N,bb}^{D}}| < 1$,
without (with) LD effects, in model III. $A_{CP}$ is restricted in the 
region bounded by solid (dashed) lines for $C_{7}^{eff} >0 $ 
($C_{7}^{eff} <0 $). Figures show that $A_{CP}$ is at the order of 
$10^{-4}$, which is a weak effect. $A_{CP}$  is larger for $C_{7}^{eff}>0$, 
as compared to the $C_{7}^{eff}<0$ case and it is possible that $A_{CP}$ 
vanishes and even take negative values for $\sin\theta>0$ and 
$C_{7}^{eff}<0$. Addition of LD effects enhances $A_{CP}$ small in amount.

\begin{figure}[htb]
\vskip -3.0truein
\centering
\epsfxsize=6.8in
\leavevmode\epsffile{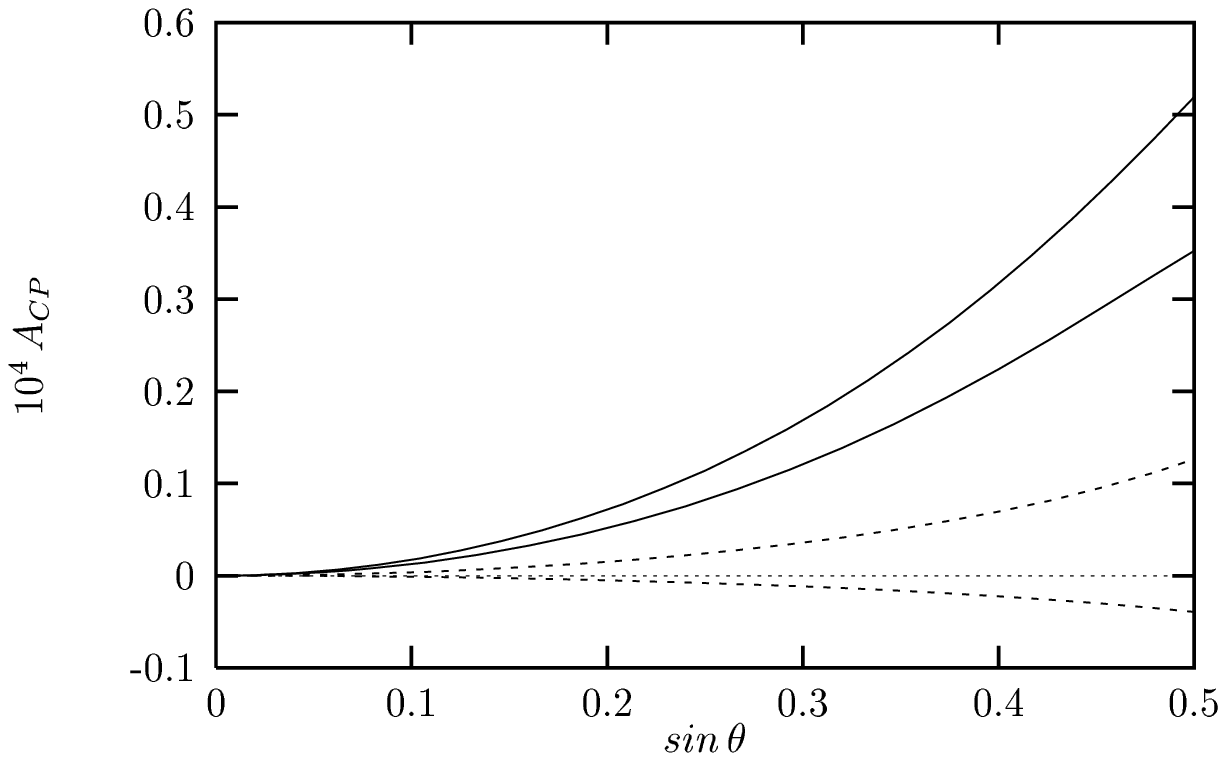}
\vskip -3.0truein
\caption[]{The same as Fig. \ref{Acp0s} but for $3HDM(O_2)$ and 
$\bar{\epsilon}^D_{N,bb}=40\, m_b$.}
\label{Acp03Hs}
\end{figure}

\begin{figure}[htb]
\vskip -3.0truein
\centering
\epsfxsize=6.8in
\leavevmode\epsffile{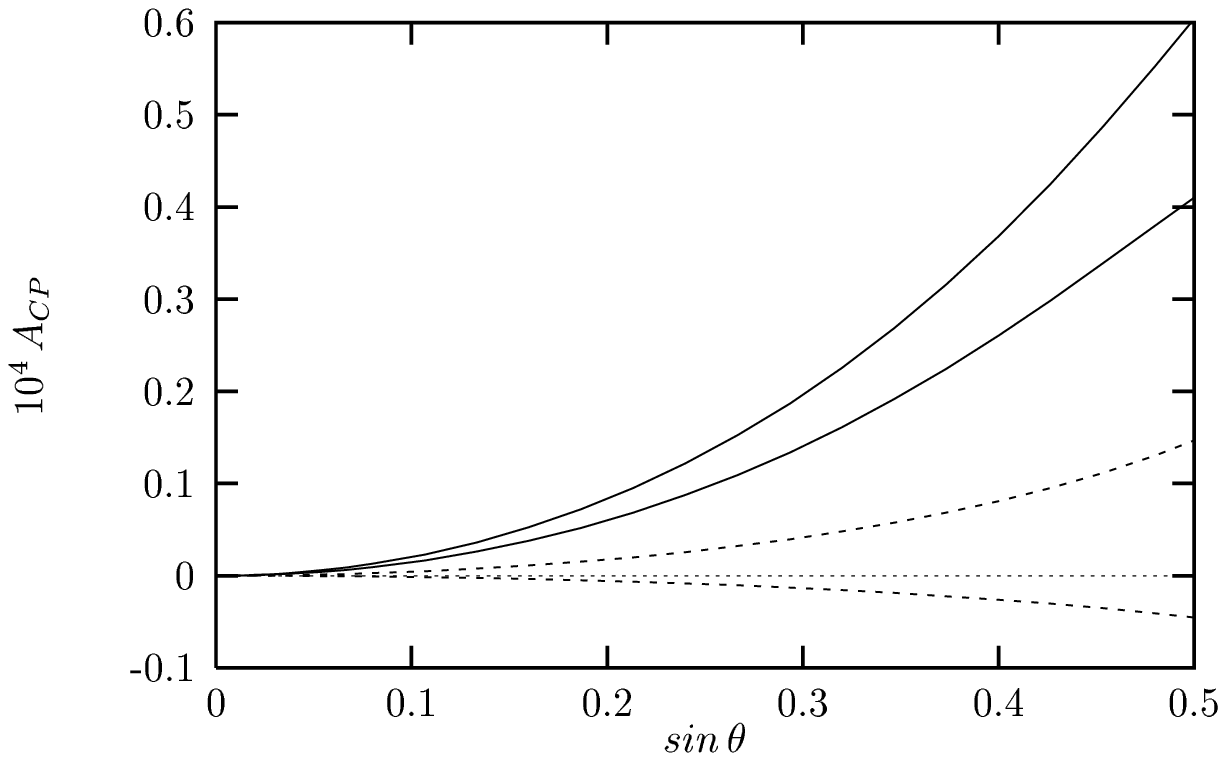}
\vskip -3.0truein
\caption[]{The same as Fig. \ref{Acp03Hs} but with LD effects.}
\label{Acpp3Hs}
\end{figure}

\begin{figure}[htb]
\vskip -3.0truein
\centering
\epsfxsize=6.8in
\leavevmode\epsffile{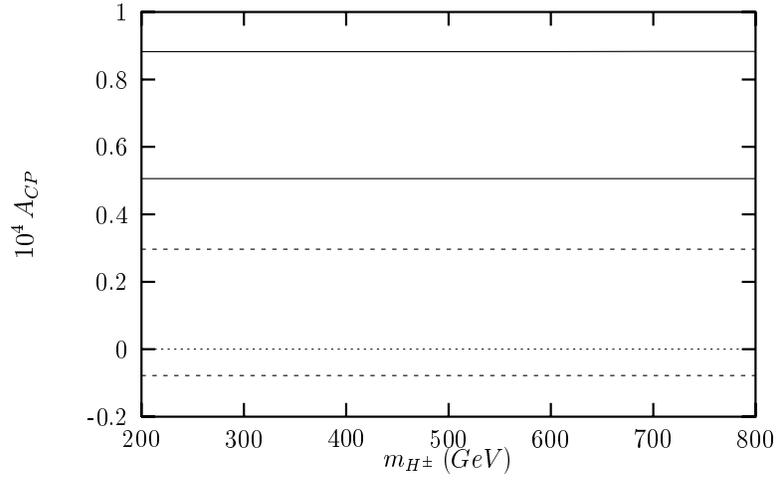}
\vskip -3.0truein
\caption[]{$A_{CP}$ as a function of $m_{H^{\pm}}$ for $sin\,\theta=0.5$ and
$\bar{\xi}_{N,bb}^{D}=40\, m_b$ in the region 
$|r_{tb}|<1$, at the scale $\mu=m_b/2$, with LD effects, in model III. Here 
$A_{CP}$ is restricted in the region bounded by solid lines for 
$C_7^{eff}>0$ and by  dashed lines for $C_7^{eff}<0$.}
\label{Acppmh}
\end{figure}

\begin{figure}[htb]
\vskip -3.0truein
\centering
\epsfxsize=6.8in
\leavevmode\epsffile{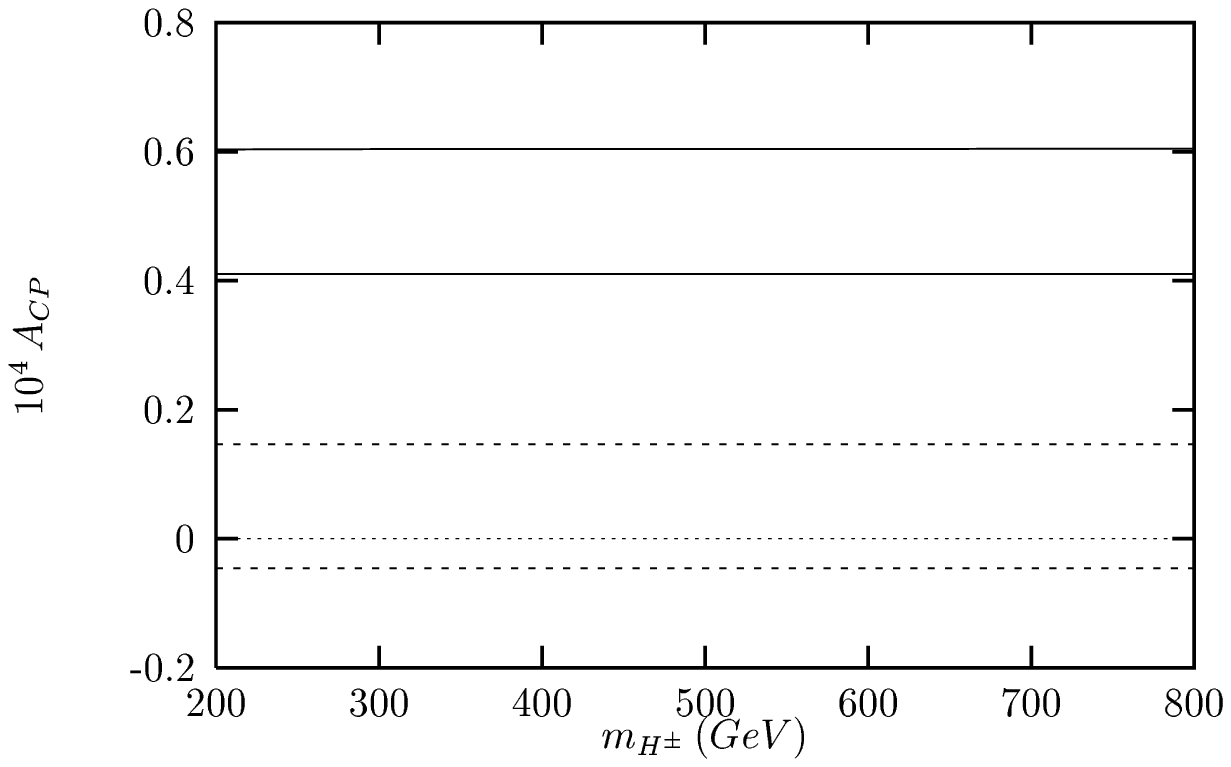}
\vskip -3.0truein
\caption[]{The same as Fig \ref{Acp0mh}, but for $3HDM(O_2)$ and 
$\bar{\epsilon}^D_{N,bb}=40\, m_b$.}
\label{Acpp3Hmh}
\end{figure}

Fig \ref{Acp03Hs}(\ref{Acpp3Hs}) denotes the same dependence of $A_{CP}$ 
for $3HDM(O_{2})$ with (without) LD effects. The area of the restricted 
region and the possible values of $A_{CP}$ are smaller as compared to the 
2HDM case. For example, for $\sin\theta=0.5$ and $C_{7}^{eff}>0$, the upper 
limit of $A_{CP}$ decreases  by an amount of $50\% $. However, the order of 
$A_{CP}$ still remains the same, namely $10^{-4}$. Furthermore, $A_{CP}$ is 
not sensitive to the charged Higgs boson mass $m_{H^{\pm}}$ 
(see Fig \ref{Acppmh} and (\ref{Acpp3Hmh}). 

\begin{figure}[htb]
\vskip -3.0truein
\centering
\epsfxsize=6.8in
\leavevmode\epsffile{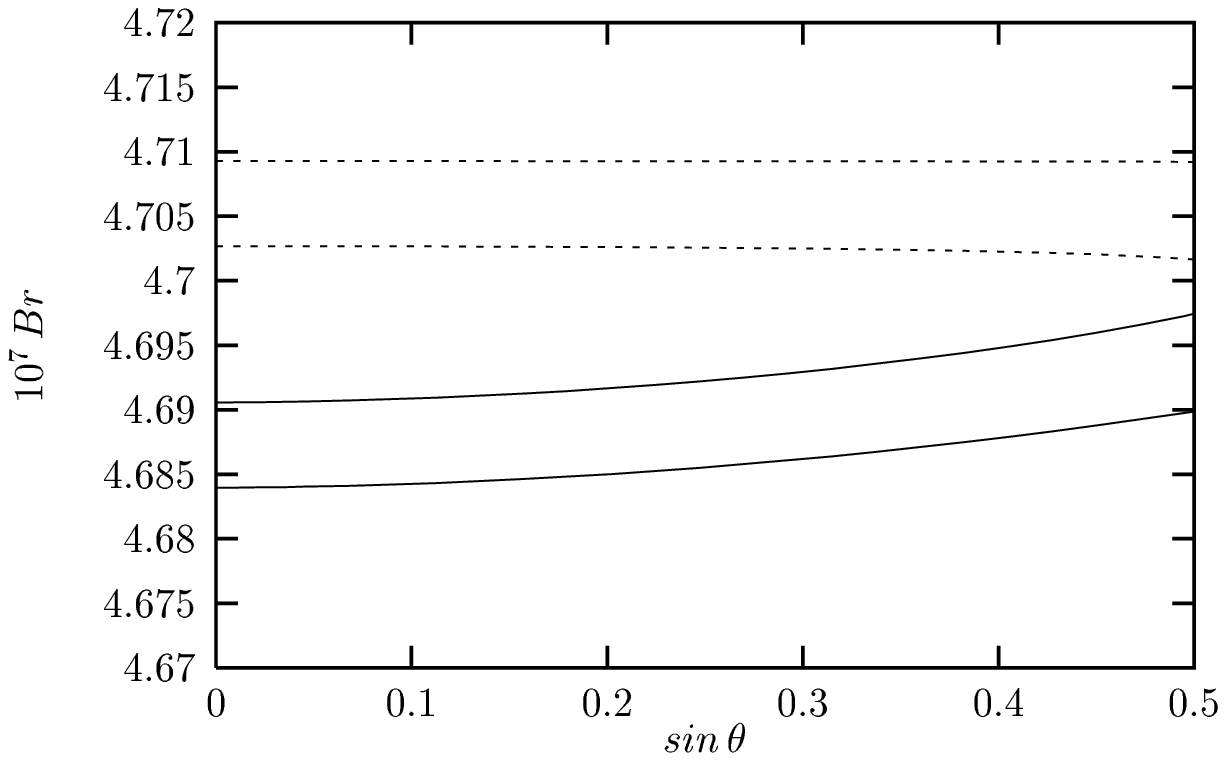}
\vskip -3.0truein
\caption[]{$Br$ as a function of  $sin\,\theta$ for
$\bar{\xi}_{N,bb}^{D}=40\, m_b$ and $m_H^{\pm}=400\, GeV$ in the region 
$|r_{tb}|<1$, at the scale $\mu=m_b/2$, without LD effects, in model III. 
Here $Br$ is restricted in the region bounded by solid lines for 
$C_7^{eff}>0$ and by  dashed lines for $C_7^{eff}<0$. }
\label{brs}
\end{figure}

\begin{figure}[htb]
\vskip -3.0truein
\centering
\epsfxsize=6.8in
\leavevmode\epsffile{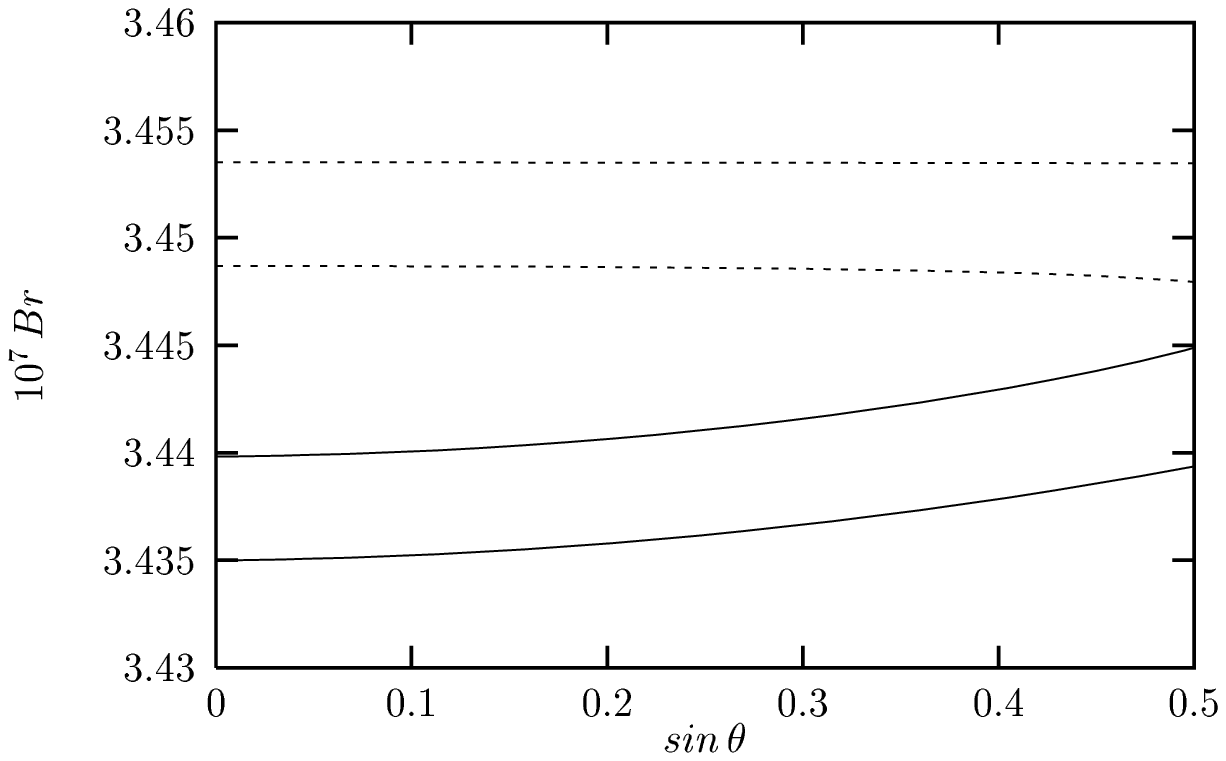}
\vskip -3.0truein
\caption[]{The same as Fig \ref{brs}, but with LD effects.}
\label{brsp}
\end{figure}

\begin{figure}[htb]
\vskip -3.0truein
\centering
\epsfxsize=6.8in
\leavevmode\epsffile{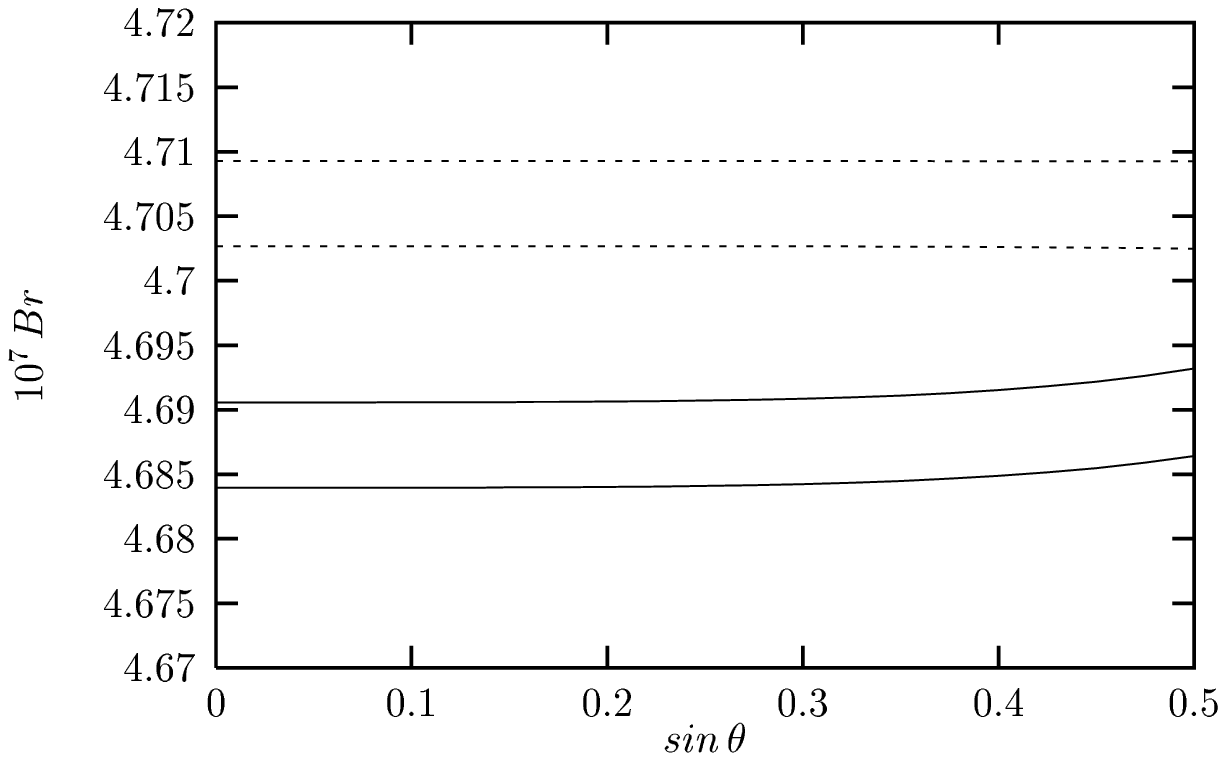}
\vskip -3.0truein
\caption[]{The same as Fig \ref{brs}, but for $3HDM(O_2)$ and 
$\bar{\epsilon}^D_{N,bb}=40\, m_b$.}
\label{brs3H}
\end{figure}

\begin{figure}[htb]
\vskip -3.0truein
\centering
\epsfxsize=6.8in
\leavevmode\epsffile{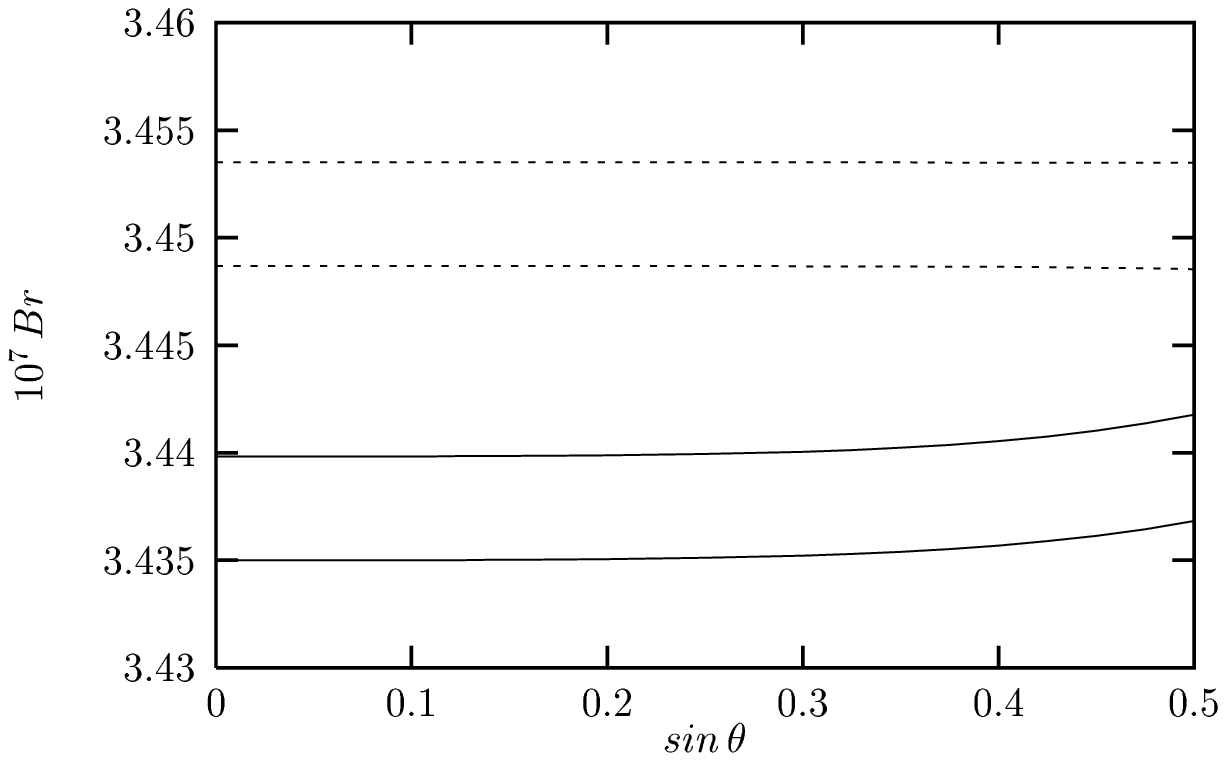}
\vskip -3.0truein
\caption[]{The same as Fig \ref{brs3H}, but with LD effects.}
\label{brsp3H}
\end{figure}

Figures \ref{brs}-\ref{brsp3H} show the $Br$ of the given process for the 
model III and $3HDM(O_2)$. In Fig \ref{brs} (\ref{brsp}), we present the 
dependence of $Br$ on $\sin\theta$ without(with) LD effects for 
$|r_{tb}| < 1$, in model III. The restricted region lies between solid 
(dashed) lines for $C_{7}^{eff}>0$ ($C_{7}^{eff}<0$). Note that $Br$ for 
$C_{7}^{eff}<0$ is greater than the one for $C_{7}^{eff}>0$. It is observed 
that the enhancement of $Br$ in the SM is negligible ($Br_{SM}=3.45\, 
10^{-7}$ with LD effects and $Br_{SM}=4.71\, 10^{-7}$ without LD effects). 
Therefore model III can not be distinguished from the SM with the 
measurement of $Br$ of the given process, for $|r_{tb}| < 1$. 
Fig \ref{brs3H} and (\ref{brsp3H} denote the same dependence for 
$3HDM(O_{2})$ and the results are similar to the 2HDM case.

\begin{figure}[htb]
\vskip -3.0truein
\centering
\epsfxsize=6.8in
\leavevmode\epsffile{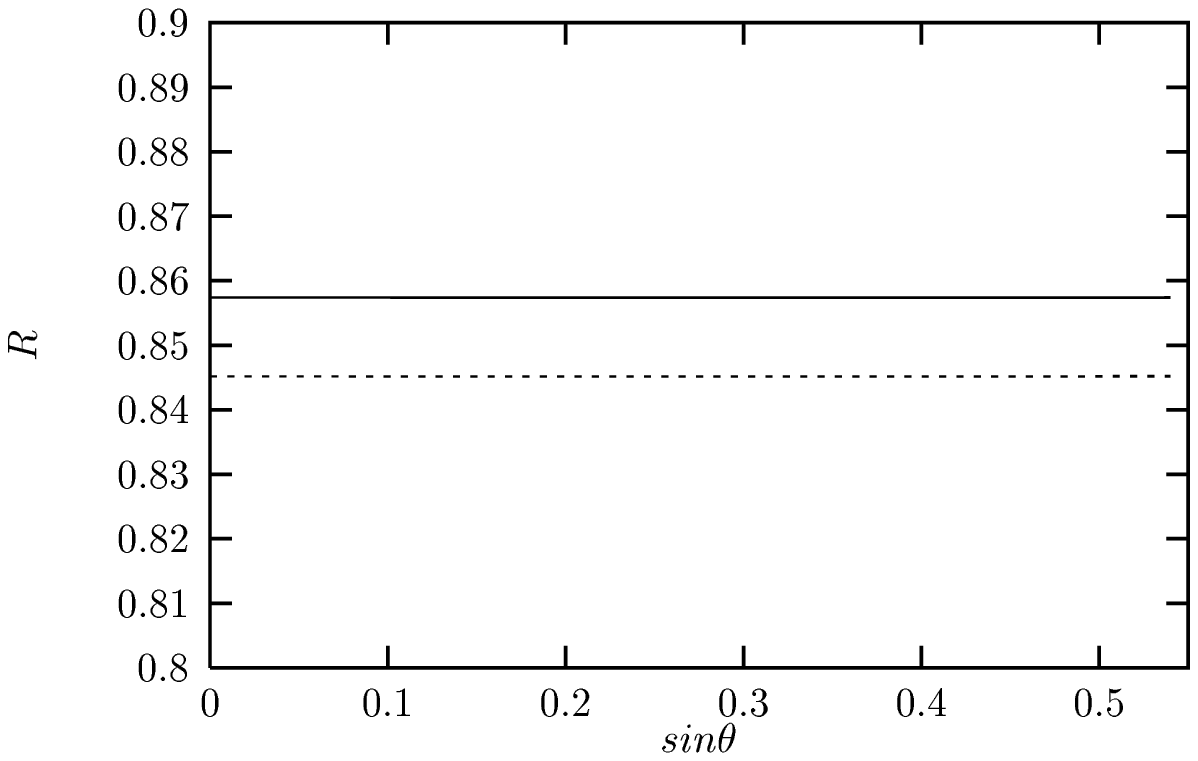}
\vskip -3.0truein
\caption[]{$R$ ratio as a function of  $sin\,\theta$ for
$\bar{\xi}_{N,bb}^{D}=40\, m_b$ and $m_H^{\pm}=400\, GeV$ in the region 
$|r_{tb}|<1$, at the scale $\mu=m_b/2$, in model III. Here $R$ ratio is 
restricted on the solid (dashed) line for $C_7^{eff}>0$ ($C_7^{eff}<0$). 
Note that, for $3HDM(O_2)$, $R$ ratio is almost the same as the one 
calculated in model III.}
\label{Rtets}
\end{figure}

Finally, in Fig \ref{Rtets}, we plot the $CP$ ratio 
$R=|\frac{ A^{+}}{A^{-}}|$ with respect to $\sin\theta$ for $2HDM$. 
Here, the solid line corresponds to $C_{7}^{eff}>0$  and the
dashed line to $C_{7}^{eff}<0$. This figure shows that $R$  is
not sensitive to $CP$ violating parameter $\theta $
and there is a small enhancement compared to the SM value ($R_{SM}=
0.845$). This ratio is also non-sensitive to the models under
consideration. 

For completeness, we would like to note that there are some 
uncertainities coming from the choice of the bound state parameters 
$m_b^{eff}$, $\bar{\Lambda}_s$ and the decay constant $f_{B_s}$. 
The physical quantities are sensitive to these parameters. For example, 
the larger $m_b^{eff}$ (smaller $\bar{\Lambda}_s$), the larger $Br$, $R$ 
and the smaller $A_{CP}$. 

In conclusion, we study  the $CP$ asymmetry of $B_{s}\rightarrow\gamma\gamma$
decay in the framework of the model III and $3HDM(O_{2})$. Further, we analyze
the $Br$ and $R$ ratio of the given process. We can summarize the main 
points of our results:

\begin{itemize}

\item In model III, a weak  $A_{CP}$ is possible and it is at the order of 
$10^{-4}$. This effect increases with the addition of LD contributions and 
this holds also in the $3HDM(O_{2})$ model. The  measurement of such a small
value of $A_{CP}$  can give information about the sign of $C_{7}^{eff}$.

\item The $Br$ ratio is not sensitive to $CP$ violation parameter $\theta$
and the enhancement as compared to SM is negligible in both models, for 
$|r_{tb}|< 1$

\item The $R$ ratio is also non-sensitive to the parameter $\theta$ in both
models.

\end{itemize}
\newpage

\begin{appendix}
\section{Appendix \\ Model III}
The Yukawa interaction for the 2HDM in the general case is
\begin{eqnarray}
{\cal{L}}_{Y}=\eta^{U}_{ij} \bar{Q}_{i L} \tilde{\phi_{1}} U_{j R}+
\eta^{D}_{ij} \bar{Q}_{i L} \phi_{1} D_{j R}+
\xi^{U}_{ij} \bar{Q}_{i L} \tilde{\phi_{2}} U_{j R}+
\xi^{D}_{ij} \bar{Q}_{i L} \phi_{2} D_{j R} + h.c. \,\,\, ,
\label{lagrangian}
\end{eqnarray}
where $L$ and $R$ denote chiral projections $L(R)=1/2(1\mp \gamma_5)$,
$\phi_{i}$ for $i=1,2$, are the two scalar doublets, $\eta^{U,D}_{ij}$
and $\xi^{U,D}_{ij}$ are the matrices of the Yukawa couplings.
With the choice of $\phi_1$ and $\phi_2$ \cite{atwood}
\begin{eqnarray}
\phi_{1}=\frac{1}{\sqrt{2}}\left[\left(\begin{array}{c c} 
0\\v+H^{0}\end{array}\right)\; + \left(\begin{array}{c c} 
\sqrt{2} \chi^{+}\\ i \chi^{0}\end{array}\right) \right]\, ; 
\phi_{2}=\frac{1}{\sqrt{2}}\left(\begin{array}{c c} 
\sqrt{2} H^{+}\\ H_1+i H_2 \end{array}\right) \,\, ,
\label{choice}
\end{eqnarray}
and the vacuum expectation values,  
\begin{eqnarray}
<\phi_{1}>=\frac{1}{\sqrt{2}}\left(\begin{array}{c c} 
0\\v\end{array}\right) \,  \, ; 
<\phi_{2}>=0 \,\, ,
\label{choice2}
\end{eqnarray}
we can write the Flavor Changing (FC) part of the interaction as 
\begin{eqnarray}
{\cal{L}}_{Y,FC}=
\xi^{U}_{ij} \bar{Q}_{i L} \tilde{\phi_{2}} U_{j R}+
\xi^{D}_{ij} \bar{Q}_{i L} \phi_{2} D_{j R} + h.c. \,\, ,
\label{lagrangianFC}
\end{eqnarray}
where the couplings  $\xi^{U,D}$ for the FC charged interactions are
\begin{eqnarray}
\xi^{U}_{ch}&=& \xi_{neutral} \,\, V_{CKM} \nonumber \,\, ,\\
\xi^{D}_{ch}&=& V_{CKM} \,\, \xi_{neutral} \,\, ,
\label{ksi1} 
\end{eqnarray}
and  $\xi^{U,D}_{neutral}$ 
\footnote{In all next discussion we denote $\xi^{U,D}_{neutral}$ 
as $\xi^{U,D}_{N}$.} 
is defined by the expression
\begin{eqnarray}
\xi^{U,D}_{N}=(V_L^{U,D})^{-1} \xi^{U,D} V_R^{U,D}\,\, .
\label{ksineut}
\end{eqnarray}
Here, the charged couplings appear as a linear combinations of neutral 
couplings multiplied by $V_{CKM}$ matrix elements.
\section{Appendix \\ 3HDM($O_2$)}
In the 3HDM the general Yukawa interaction is, 
\begin{eqnarray}
{\cal{L}}_{Y}&=&\eta^{U}_{ij} \bar{Q}_{i L} \tilde{\phi_{1}} U_{j R}+
\eta^{D}_{ij} \bar{Q}_{i L} \phi_{1} D_{j R}+
\xi^{U}_{ij} \bar{Q}_{i L} \tilde{\phi_{2}} U_{j R}+
\xi^{D}_{ij} \bar{Q}_{i L} \phi_{2} D_{j R} \nonumber \\
&+&
\rho^{U}_{ij} \bar{Q}_{i L} \tilde{\phi_{3}} U_{j R}+
\rho^{D}_{ij} \bar{Q}_{i L} \phi_{3} D_{j R}
 + h.c. \,\,\, ,
\label{lagrangian3H}
\end{eqnarray}
where $\phi_{i}$ for $i=1,2,3$, are three scalar doublets and  
$\eta^{U,D}_{ij}$, $\xi^{U,D}_{ij}$, $\rho^{U,D}_{ij}$ are
the Yukawa matrices having complex entries, in general. Now, we choose 
scalar Higgs doublets such that the first one describes only the SM part 
and last two carry the information about new physics beyond the SM: 
\begin{eqnarray}
\phi_{1}=\frac{1}{\sqrt{2}}\left[\left(\begin{array}{c c} 
0\\v+H^{0}\end{array}\right)\; + \left(\begin{array}{c c} 
\sqrt{2} \chi^{+}\\ i \chi^{0}\end{array}\right) \right]\, , 
\nonumber \\ \\
\phi_{2}=\frac{1}{\sqrt{2}}\left(\begin{array}{c c} 
\sqrt{2} H^{+}\\ H^1+i H^2 \end{array}\right) \,\, ,\,\, 
\phi_{3}=\frac{1}{\sqrt{2}}\left(\begin{array}{c c} 
\sqrt{2} F^{+}\\ H^3+i H^4 \end{array}\right) \,\, ,\nonumber
\label{choice3H}
\end{eqnarray}
with the vacuum expectation values,  
\begin{eqnarray}
<\phi_{1}>=\frac{1}{\sqrt{2}}\left(\begin{array}{c c} 
0\\v\end{array}\right) \,  \, ; 
<\phi_{2}>=0 \,\, ; <\phi_{3}>=0\,\, . 
\label{cb}
\end{eqnarray}
Note that, the similar choice was done in the literature for the general 
2HDM (model III)  \cite{atwood}. The Yukawa interaction responsible for the 
Flavor Changing (FC) interactions is 
\begin{eqnarray}
{\cal{L}}_{Y,FC}=
\xi^{U}_{ij} \bar{Q}_{i L} \tilde{\phi_{2}} U_{j R}+
\xi^{D}_{ij} \bar{Q}_{i L} \phi_{2} D_{j R}
+\rho^{U}_{ij} \bar{Q}_{i L} \tilde{\phi_{3}} U_{j R}+
\rho^{D}_{ij} \bar{Q}_{i L} \phi_{3} D_{j R} + h.c. \,\, .
\label{lagrangianFC3H}
\end{eqnarray}
and the couplings  $\xi^{U,D}$ and $\rho^{U,D}$ for the charged FC 
interactions are 
\begin{eqnarray}
\xi^{U}_{ch}&=& \xi_{N} \,\, V_{CKM} \nonumber \,\, ,\\
\xi^{D}_{ch}&=& V_{CKM} \,\, \xi_{N}  \nonumber \,\, , \\
\rho^{U}_{ch}&=& \rho_{N} \,\, V_{CKM} \nonumber \,\, ,\\
\rho^{D}_{ch}&=& V_{CKM} \,\, \rho_{N} \,\, ,
\label{ksi13H} 
\end{eqnarray}
and
\begin{eqnarray}
\xi^{U,D}_{N}&=&(V_L^{U,D})^{-1} \xi^{U,D}\, V_R^{U,D}\,\, , \nonumber \\
\rho^{U,D}_{N}&=&(V_L^{U,D})^{-1} \rho^{U,D}\, V_R^{U,D}\,\, ,
\label{ksineut3H}
\end{eqnarray}

In the 3HDM model, the Higgs sector is extended and therefore the number of 
free parameters, namely, masses of new charged and neutral 
Higgs particles, new Yukawa couplings, increases.  Fortunately, by introducing 
a new transformation in the Higgs sector and taking the 3HDM Lagrangian 
invariant under it, the number of free parameters can be reduced enormously 
\cite{eril5}. Taking the following $O(2)$ transformation: 
\begin{eqnarray}
\phi'_{1}&=&\phi_{1}\nonumber \,\,,\\
\phi'_{2}&=&cos\,\alpha\,\, \phi_2+sin\,\alpha\,\, \phi_3 \,\, , \nonumber \\
\phi'_{3}&=&-sin\,\alpha\,\, \phi_2 + cos\,\alpha\,\, \phi_3\,\,,
\label{trans}
\end{eqnarray}
where $\alpha$ is the global parameter, which represents a rotation of 
the vectors $\phi_2$ and $\phi_3$ along the axis where $\phi_1$ lies 
and assuming the invariance of the gauge and $CP$ invariant Higgs 
potential 
\begin{eqnarray}
V(\phi_1, \phi_2,\phi_3 )&=&c_1 (\phi_1^+ \phi_1-v^2/2)^2+
c_2 (\phi_2^+ \phi_2)^2 \nonumber \\ &+& 
c_3 (\phi_3^+ \phi_3)^2+
c_4 [(\phi_1^+ \phi_1-v^2/2)+ \phi_2^+ \phi_2+\phi_3^+ \phi_3]^2
\nonumber \\ &+& 
c_5 [(\phi_1^+ \phi_1) (\phi_2^+ \phi_2)-(\phi_1^+ \phi_2)(\phi_2^+ \phi_1)]
\nonumber \\ &+& 
c_6 [(\phi_1^+ \phi_1) (\phi_3^+ \phi_3)-(\phi_1^+ \phi_3)(\phi_3^+ \phi_1)]
\nonumber \\ &+& 
c_7 [(\phi_2^+ \phi_2) (\phi_3^+ \phi_3)-(\phi_2^+ \phi_3)(\phi_3^+ \phi_2)]
\nonumber \\ &+& 
c_8 [Re(\phi_1^+ \phi_2)]^2 +c_9 [Re(\phi_1^+ \phi_3)]^2 +
c_{10} [Re(\phi_2^+ \phi_3)]^2 \nonumber \\ &+&
c_{11} [Im(\phi_1^+ \phi_2)]^2 +c_{12} [Im(\phi_1^+ \phi_3)]^2 +
c_{13} [Im(\phi_2^+ \phi_3)]^2 +c_{14}
\label{potential}
\end{eqnarray}
we get the masses of new particles as
\begin{eqnarray}
m_{F^\pm}=m_{H^\pm}\nonumber \,\, , \\
m_{H^3}=m_{H^1}\nonumber \,\, , \\
m_{H^4}=m_{H^2}\nonumber \,\, , \\
\label{mass2}
\end{eqnarray}
Further, the application of this transformation to the Yukawa Lagrangian 
eq.(\ref{lagrangian3H}) allows us to write the equality 
\begin{eqnarray}
(\xi^{\prime U(D)})^+ \xi^{\prime U (D) } +
(\rho^{\prime U (D)})^+\rho^{\prime U (D) }=
(\xi^{U (D)})^+ \xi^{U (D)} +
(\rho^{U (D)})^+ \rho^{U (D) }\,\, , 
\label{yukinv}
\end{eqnarray}
where
\begin{eqnarray}
\xi^{\prime U(D)}_{ij}&=& \xi^{U (D)}_{ij} cos\, \alpha+
\rho^{U(D)}_{ij} sin\, \alpha\,\, \nonumber ,\\
\rho^{\prime U (D)}_{ij}&=&-\xi^{U (D)}_{ij} sin\, \alpha+
\rho^{U (D)}_{ij} cos\, \alpha \,\, .
\label{yuktr}
\end{eqnarray}
and therefore the Yukawa matrices $\xi^{U(D)}$ and $\rho^{U(D)}$
can be parametrized as,  
\begin{eqnarray}
\xi^{U (D)}=\epsilon^{U(D)} cos\,\theta \nonumber \,\, ,\\
\rho^{U}=\epsilon^{U} sin\,\theta \nonumber \,\, ,\\ 
\rho^{D}=i \epsilon^{D} sin\,\theta \,\, .
\label{yukpar}
\end{eqnarray}
Here $\epsilon^{U(D)}$ are real matrices satisfy the equation 
\begin{eqnarray}
(\xi^{\prime U(D)})^+ \xi^{\prime U (D) } +
(\rho^{\prime U (D)})^+ \rho^{\prime U (D) }=
(\epsilon^{U(D)})^T \epsilon^{U(D)} 
\label{yukpareq}
\end{eqnarray}
and $T$ denotes transpose operation. 
Finally, we could reduce the number of the Yukawa matrices  $\xi^{U,(D)}$ 
and $\rho^{U,(D)}$, by connecting them with the expression given in 
eq.(\ref{yukpar}). Further, we take into account
only the Yukawa couplings $\xi^{U}_{N, tt}$, $\xi^{D}_{N,bb}$, 
$\rho^{U}_{N,tt}$ and $\rho^{D}_{N,bb}$, since we assume that 
the others are small due to the discussion given in \cite{alil}.
\section{Appendix \\ The operator basis and the Wilson coefficients 
for the decay $b\rightarrow s \gamma\gamma$ in the model III}
The operator basis is the same as the one used for the $b\rightarrow s
\gamma$ decay in the  model III \cite{alil} and 
$SU(2)_L\times SU(2)_R\times U(1)$ extensions of the SM \cite{cho}:
\begin{eqnarray}
 O_1 &=& (\bar{s}_{L \alpha} \gamma_\mu c_{L \beta})
               (\bar{c}_{L \beta} \gamma^\mu b_{L \alpha}), \nonumber   \\
 O_2 &=& (\bar{s}_{L \alpha} \gamma_\mu c_{L \alpha})
               (\bar{c}_{L \beta} \gamma^\mu b_{L \beta}),  \nonumber   \\
 O_3 &=& (\bar{s}_{L \alpha} \gamma_\mu b_{L \alpha})
               \sum_{q=u,d,s,c,b}
               (\bar{q}_{L \beta} \gamma^\mu q_{L \beta}),  \nonumber   \\
 O_4 &=& (\bar{s}_{L \alpha} \gamma_\mu b_{L \beta})
                \sum_{q=u,d,s,c,b}
               (\bar{q}_{L \beta} \gamma^\mu q_{L \alpha}),   \nonumber  \\
 O_5 &=& (\bar{s}_{L \alpha} \gamma_\mu b_{L \alpha})
               \sum_{q=u,d,s,c,b}
               (\bar{q}_{R \beta} \gamma^\mu q_{R \beta}),   \nonumber  \\
 O_6 &=& (\bar{s}_{L \alpha} \gamma_\mu b_{L \beta})
                \sum_{q=u,d,s,c,b}
               (\bar{q}_{R \beta} \gamma^\mu q_{R \alpha}),  \nonumber   \\  
 O_7 &=& \frac{e}{16 \pi^2}
          \bar{s}_{\alpha} \sigma_{\mu \nu} (m_b R + m_s L) b_{\alpha}
                {\cal{F}}^{\mu \nu},                             \nonumber       \\
 O_8 &=& \frac{g}{16 \pi^2}
    \bar{s}_{\alpha} T_{\alpha \beta}^a \sigma_{\mu \nu} (m_b R + m_s L)  
          b_{\beta} {\cal{G}}^{a \mu \nu} \nonumber \,\, , \\  
 O_9 &=& (\bar{s}_{L \alpha} \gamma_\mu c_{L \beta})
               (\bar{c}_{R \beta} \gamma^\mu b_{R \alpha}), \nonumber   \\
 O_{10} &=& (\bar{s}_{L \alpha} \gamma_\mu c_{L \alpha})
(\bar{c}_{R \beta} \gamma^\mu b_{R \beta}),
\label{op1}
\end{eqnarray}
and the second operator set $O'_{1} - O'_{10}$ which are 
flipped chirality partners of $O_{1} - O_{10}$:
\begin{eqnarray}
 O'_1 &=& (\bar{s}_{R \alpha} \gamma_\mu c_{R \beta})
               (\bar{c}_{R \beta} \gamma^\mu b_{R \alpha}), \nonumber   \\
 O'_2 &=& (\bar{s}_{R \alpha} \gamma_\mu c_{R \alpha})
               (\bar{c}_{R \beta} \gamma^\mu b_{R \beta}),  \nonumber   \\
 O'_3 &=& (\bar{s}_{R \alpha} \gamma_\mu b_{R \alpha})
               \sum_{q=u,d,s,c,b}
               (\bar{q}_{R \beta} \gamma^\mu q_{R \beta}),  \nonumber   \\
 O'_4 &=& (\bar{s}_{R \alpha} \gamma_\mu b_{R \beta})
                \sum_{q=u,d,s,c,b}
               (\bar{q}_{R \beta} \gamma^\mu q_{R \alpha}),   \nonumber  \\
 O'_5 &=& (\bar{s}_{R \alpha} \gamma_\mu b_{R \alpha})
               \sum_{q=u,d,s,c,b}
               (\bar{q}_{L \beta} \gamma^\mu q_{L \beta}),   \nonumber  \\
 O'_6 &=& (\bar{s}_{R \alpha} \gamma_\mu b_{R \beta})
                \sum_{q=u,d,s,c,b}
               (\bar{q}_{L \beta} \gamma^\mu q_{L \alpha}),  \nonumber   \\  
 O'_7 &=& \frac{e}{16 \pi^2}
          \bar{s}_{\alpha} \sigma_{\mu \nu} (m_b L + m_s R) b_{\alpha}
                {\cal{F}}^{\mu \nu},                             \nonumber       \\
 O'_8 &=& \frac{g}{16 \pi^2}
    \bar{s}_{\alpha} T_{\alpha \beta}^a \sigma_{\mu \nu} (m_b L + m_s R)  
          b_{\beta} {\cal{G}}^{a \mu \nu}, \nonumber \\ 
 O'_9 &=& (\bar{s}_{R \alpha} \gamma_\mu c_{R \beta})
               (\bar{c}_{L \beta} \gamma^\mu b_{L \alpha})\,\, , \nonumber   \\
 O'_{10} &=& (\bar{s}_{R \alpha} \gamma_\mu c_{R \alpha})
(\bar{c}_{L \beta} \gamma^\mu b_{L \beta})\,\, ,
\label{op2}
\end{eqnarray}
where  
$\alpha$ and $\beta$ are $SU(3)$ colour indices and
${\cal{F}}^{\mu \nu}$ and ${\cal{G}}^{\mu \nu}$
are the field strength tensors of the electromagnetic and strong
interactions, respectively. In the calculations, we take only the charged 
Higgs contributions into account and neglect the effects of neutral Higgs 
bosons (see \cite{alil4} for details).
Further, in our expressions we use the redefinition,
\begin{eqnarray}
\xi^{U,D}=\sqrt\frac{4 G_F}{\sqrt 2}\, \bar{\xi}^{U,D}\,\,.
\label{xi}
\end{eqnarray}

Denoting the Wilson coefficients for the SM with $C_{i}^{SM}(m_{W})$ and the
additional charged Higgs contribution with $C_{i}^{H}(m_{W})$, 
we have the initial values for the first set of operators 
(eq.(~\ref{op1})) (\cite{alil} and references within) 
\begin{eqnarray}
C^{SM}_{1,3,\dots 6,9,10}(m_W)&=&0 \nonumber \, \, , \\
C^{SM}_2(m_W)&=&1 \nonumber \, \, , \\
C_7^{SM}(m_W)&=&\frac{3 x^3-2 x^2}{4(x-1)^4} \ln x+
\frac{-8x^3-5 x^2+7 x}{24 (x-1)^3} \nonumber \, \, , \\
C_8^{SM}(m_W)&=&-\frac{3 x^2}{4(x-1)^4} \ln x+
\frac{-x^3+5 x^2+2 x}{8 (x-1)^3}\nonumber \, \, , \\ 
C^{H}_{1,\dots 6,9,10}(m_W)&=&0 \nonumber \, \, , \\
C_7^{H}(m_W)&=&\frac{1}{m_{t}^2} \,
(\bar{\xi}^{U}_{N,tt}+\bar{\xi}^{U}_{N,tc}
\frac{V_{cs}^{*}}{V_{ts}^{*}}) \, (\bar{\xi}^{U}_{N,tt}+\bar{\xi}^{U}_{N,tc}
\frac{V_{cb}}{V_{tb}}) F_{1}(y)\nonumber  \, \, , \\
&+&\frac{1}{m_t m_b} \, (\bar{\xi}^{U}_{N,tt}+\bar{\xi}^{U}_{N,tc}
\frac{V_{cs}^{*}}{V_{ts}^{*}}) \, (\bar{\xi}^{D}_{N,bb}+\bar{\xi}^{D}_{N,sb}
\frac{V_{ts}}{V_{tb}}) F_{2}(y)\nonumber  \, \, , \\
C_8^{H}(m_W)&=&\frac{1}{m_{t}^2} \,
(\bar{\xi}^{U}_{N,tt}+\bar{\xi}^{U}_{N,tc}
\frac{V_{cs}^{*}}{V_{ts}^{*}}) \, (\bar{\xi}^{U}_{N,tt}+\bar{\xi}^{U}_{N,tc}
\frac{V_{cb}}{V_{tb}})G_{1}(y)\nonumber  \, \, , \\
&+&\frac{1}{m_t m_b} \, (\bar{\xi}^{U}_{N,tt}+\bar{\xi}^{U}_{N,tc}
\frac{V_{cs}^{*}}{V_{ts}^{*}}) \, (\bar{\xi}^{D}_{N,bb}+\bar{\xi}^{U}_{N,sb}
\frac{V_{ts}}{V_{tb}}) G_{2}(y) \, \, ,
\label{CoeffH}
\end{eqnarray}
and for the second set of operators eq.~(\ref{op2}), 
\begin{eqnarray}
C^{\prime SM}_{1,\dots 10}(m_W)&=&0 \nonumber \, \, , \\
C^{\prime H}_{1,\dots 6,9,10}(m_W)&=&0 \nonumber \, \, , \\
C^{\prime H}_7(m_W)&=&\frac{1}{m_t^2} \,
(\bar{\xi}^{D}_{N,bs}\frac{V_{tb}}{V_{ts}^{*}}+\bar{\xi}^{D}_{N,ss})
\, (\bar{\xi}^{D}_{N,bb}+\bar{\xi}^{D}_{N,sb}
\frac{V_{ts}}{V_{tb}}) F_{1}(y)\nonumber  \, \, , \\
&+& \frac{1}{m_t m_b}\, (\bar{\xi}^{D}_{N,bs}\frac{V_{tb}}{V_{ts}^{*}}
+\bar{\xi}^{D}_{N,ss}) \, (\bar{\xi}^{U}_{N,tt}+\bar{\xi}^{U}_{N,tc}
\frac{V_{cb}}{V_{tb}}) F_{2}(y)\nonumber  \, \, , \\
C^{\prime H}_8 (m_W)&=&\frac{1}{m_t^2} \,
(\bar{\xi}^{D}_{N,bs}\frac{V_{tb}}{V_{ts}^{*}}+\bar{\xi}^{D}_{N,ss})
\, (\bar{\xi}^{D}_{N,bb}+\bar{\xi}^{D}_{N,sb}
\frac{V_{ts}}{V_{tb}}) G_{1}(y)\nonumber  \, \, , \\
&+&\frac{1}{m_t m_b} \, (\bar{\xi}^{D}_{N,bs}\frac{V_{tb}}{V_{ts}^{*}}
+\bar{\xi}^{D}_{N,ss}) \, (\bar{\xi}^{U}_{N,tt}+\bar{\xi}^{U}_{N,tc}
\frac{V_{cb}}{V_{tb}}) G_{2}(y) \,\, ,
\label{CoeffH2}
\end{eqnarray}
where $x=m_t^2/m_W^2$ and $y=m_t^2/m_{H^{\pm}}^2$.
The functions $F_{1}(y)$, $F_{2}(y)$, $G_{1}(y)$ and $G_{2}(y)$ are given as
\begin{eqnarray}
F_{1}(y)&=& \frac{y(7-5y-8y^2)}{72 (y-1)^3}+\frac{y^2 (3y-2)}{12(y-1)^4}
\,
ln y \nonumber  \,\, , \\ 
F_{2}(y)&=& \frac{y(5y-3)}{12 (y-1)^2}+\frac{y(-3y+2)}{6(y-1)^3}\, ln y 
\nonumber  \,\, ,\\ 
G_{1}(y)&=& \frac{y(-y^2+5y+2)}{24 (y-1)^3}+\frac{-y^2} {4(y-1)^4} \, ln y
\nonumber  \,\, ,\\ 
G_{2}(y)&=& \frac{y(y-3)}{4 (y-1)^2}+\frac{y} {2(y-1)^3} \, ln y \,\, .
\label{F1G1}
\end{eqnarray}
Note that we neglect the contributions of the internal $u$ and $c$ quarks 
compared to one due to the internal $t$ quark. In the 
$3HDM(O_2)$ model, the Wilson coefficients are obtained with the 
replacement $\bar{\xi}^{U(D)}\rightarrow \bar{\epsilon}^{U(D)}$ and
the redefinition of $\lambda_{\theta}$, 
$\lambda_{\theta}=\frac{1}{m_t m_b} \bar{\epsilon}^{U}_{N,tt}
 \bar{\epsilon}^{D}_{N,bb} (cos^2\,\theta+ i \, sin^2\,\theta$).

For the initial values of the Wilson coefficients in the model III  
(eqs. (\ref{CoeffH})and (\ref{CoeffH2})), we have 
\begin{eqnarray}
C^{2HDM}_{1,3,\dots 6,9,10}(m_W)&=&0 \nonumber \, \, , \\
C_2^{2HDM}(m_W)&=&1 \nonumber \, \, , \\
C_7^{2HDM}(m_W)&=&C_7^{SM}(m_W)+C_7^{H}(m_W) \nonumber \, \, , \\
C_8^{2HDM}(m_W)&=&C_8^{SM}(m_W)+C_8^{H}(m_W) \nonumber \, \, , \\ 
C^{\prime 2HDM}_{1,2,3,\dots 6,9,10}(m_W)&=&0 \nonumber \, \, , \\
C_7^{\prime 2HDM}(m_W)&=&C_7^{\prime SM}(m_W)+C_7^{\prime H}(m_W) \nonumber \, \, , \\
C_8^{\prime 2HDM}(m_W)&=&C_8^{\prime SM}(m_W)+C_8^{\prime H}(m_W) \, \, . 
\label{Coef2HDM}
\end{eqnarray}

At this stage it is possible to obtain the result for model II, in the
approximation $\frac{m_{s}}{m_{b}}\sim 0$ and 
$\frac{m_{b}^2}{m_{t}^2}\sim 0$, by making the
following replacements in the Wilson coefficients:
\begin{eqnarray}
\bar{\xi}^{U *}_{st}\bar{\xi}^{U}_{tb}&=&m_{t}^2
\frac{1}{tan^{2}\beta}\nonumber \,\, ,\\
\bar{\xi}^{U *}_{st}\bar{\xi}^{D}_{tb}&=&-m_{t} m_{b}\,\, ,
\label{replacement}
\end{eqnarray}
and taking zero for the coefficients of the flipped operator set, i.e
$C^{\prime}_{i}\rightarrow 0$. 

The evaluation of the Wilson coefficients are done by using the initial 
values $C_i^{2HDM\,(3HDM (O_2)}$ ($C_i^{\prime 2HDM\, (3HDM (O_2))}$) 
and  their contributions at any lower scale can be calculated as in 
the SM case 
\cite{alil}. 
\section{Appendix \\ The necessary functions used in the Wilson
coefficients}
The explicit forms of the functions $I(m_q)$, $J(m_q)$ and $\triangle(m_q)$ 
appearing in eqs. \ref{Amplitudes1} and \ref{Amplitudes2} are
\begin{eqnarray}
I(m_q)&=&1+\frac{m_q^2}{m_{B_s}^2} \triangle (m_q) \, \, , \nonumber \\
J(m_q)&=&1-\frac{m_{B_s}^2-4 m_q^2}{4 m_{B_s}^2} \triangle(m_q)  \, \, ,
\nonumber \\
\triangle(m_q)&=&\left(
\ln(\frac{m_{B_s}+\sqrt{m_{B_s}^2-4 m_q^2}}
         {m_{B_s}-\sqrt{m_{B_s}^2-4 m_q^2}})-i \pi \right)^2 
\, \,{\mbox{for}}\, \, \frac{m^2_{B_s}}{4 m_q^2} \geq 1 , \nonumber \\
\triangle(m_q)&=&-\left(
2 \arctan(\frac{\sqrt{4 m_q^2-m_{B_s}^2}}
         {m_{B_s}})-\pi \right)^2 
\, \,{\mbox{for}}\, \, \frac{m^2_{B_s}}{4 m_q^2} <\ 1 .
\end{eqnarray}

\end{appendix}

\newpage

\newpage
\begin{thebibliography}{1}
\bibitem{Hewett} J. L. Hewett, in proc. of the $21^{st}$ Annual SLAC Summer 
Institute, ed. L. De Porcel and C. Dunwoode, SLAC-PUB6521.
\bibitem{kalinowski}
        S. Herrlich and J. Kalinowski , Nucl. Phys. B {\bf 381} (1992) 501.
%
\bibitem{Ricciardi} L. Reina, G. Ricciardi and A. Soni,Phys. Lett B {\bf 396} 
(1997) 231.
%
\bibitem{yao} G.-L. Lin, J. Liu and Y.-P. Yao, Phys. Rev. Lett. {\bf 64} (1990) 1498; \\
   G.-L. Lin, J. Liu and Y.-P. Yao, Phys. Rev. D {\bf 42} (1990) 2314.
%
\bibitem{simma}
        H. Simma and D. Wyler, Nucl. Phys. B {\bf 344} (1990) 283.
%
\bibitem{aliev}
        T. M. Aliev and G. Turan, Phys. Rev. D {\bf 48} (1993) 1176.
%
\bibitem{Grinstein} B. Grinstein, R. Springer, and M. Wise,
\np\ B{\bf 339} (1990) 269; R. Grigjanis, P.J. O'Donnel,
M. Sutherland and H. Navelet, \pl\ B{\bf 213} (1988) 355; 
\pl\ B{\bf 286} (1992) E, 413; G. Cella, G. Curci, G. Ricciardi and 
A. Vicer\'e, \pl\ B{\bf 325} (1994) 227, \np\ B{\bf 431} (1994) 417; 
M. Misiak, Nucl. Phys B{\bf 393} (1993) 23, Erratum B{\bf 439} (1995) 
461.
%
\bibitem{misiak} K. G. Chetyrkin, M. Misiak and M. M\"unz, 
Phys. Lett.B  {\bf 400} (1997) 206;  
C. Greub, T. Hurth and D. Wyler, Phys. Lett.B  {\bf 380} (1996) 385;
 Phys. Rev. D {\bf 54} (1996) 3350. 
%
\bibitem{bsglocoeff} M. Ciuchini, E. Franco, G. Martinelli, L. Reina 
and L. Silvestrini, \pl\ B{\bf 316} (1993) 127; \np\ B{\bf 421} (1994) 41.
%
\bibitem{burasmisiak}
A. J. Buras, M. Misiak, M. M\"unz and S. Pokorski, \np\ B{\bf 424}
(1994) 374.
%
\bibitem{Gud}
        G. Hiller and E. Iltan, Phys. Lett. B{\bf 409} (1997) 425.
%
\bibitem{YaoLin} C. H. V. Chang, G. L. Lin and Y. P. Yao,
Phys. Lett. B{\bf 415} (1997) 395. 
%
\bibitem{Soni} L. Reina, G. Ricciardi and A. Soni, 
Phys. Rev. D{\bf 56} (1997) 5085.
%
\bibitem{gudalil}
 T. M. Aliev, G. Hiller and E. O. Iltan, Nucl. Phys. B{\bf 515} (1998) 321.
%
\bibitem{bertolini} S. Bertolini and J. Matias, Phys. Rev. D {\bf 57} (1998)
4197. 
%
\bibitem{alil4}
T. M. Aliev and E. O. Iltan, Phys. Rev. D{\bf 58 } (1998) 095014 .
%
\bibitem{Acciarri}
M. Acciarri et al. (L3 Collaboration), Phys. Lett. B {\bf 363} (1995) 127.
%
\bibitem{eril5}  E. Iltan, to appear in Phys. Rev. D, hep-ph/9903202. 
%
\bibitem{ciuchini} M. Ciuchini et al., Nucl. Phys. B{\bf 527} (1998) 21. 
%
\bibitem{alil} T. M. Aliev, and E. Iltan, J. Phys. G{\bf 25} (1999) 989,
%
\bibitem{eril4}  E. Iltan, Phys. Rev. D{\bf 60 } (1999) 034023. 
%
\bibitem{buras2}
A. J. Buras and M. M\"unz, Phys. Rev. D{\bf 52 } (1995) 186. 
%
\bibitem{gudil} G. Hiller and E. O. Iltan, Mod. Phys. Lett. A {\bf 12} (1997)
2837.
%
\bibitem{cleo} M. S. Alam et al., CLEO Collaboration, ICHEP 98 
Conference,1998 
%
\bibitem{atwood} D. Atwood, L. Reina and A. Soni, Phys. Rev. D{\bf 55} 
(1997) 3156.
%
\bibitem{Chao} D. B. Chao, K. Cheung and Keung, hep-ph/98011235 (1998)
%
\bibitem{cho} P. Cho and Misiak, Phys. Rev. D {\bf 49} (1994) 5894.\\
%
\end{thebibliography}
\end{document}